\documentclass[preprint2]{aastex}

\shorttitle{The X-ray transfer function in AGN}
\shortauthors{Legg et~al.}

\begin{document}

\title{
Direct Measurement of the X-ray Time-Delay Transfer Function in Active Galactic Nuclei.
}

\author{E.~Legg, L.~Miller}
\affil{Department of Physics, Oxford University, Keble Road, Oxford OX1 3RH, UK}

\author{T.J.~Turner, M.~Giustini}
\affil{Department of Physics, University of Maryland Baltimore County, Baltimore MD 21250, USA }

\author{J.N.~Reeves}
\affil{Astrophysics Group, School of Physical and Geographical Sciences, Keele University, Keele, Staffordshire ST5 5BG, UK}
\affil{Department of Physics, University of Maryland Baltimore County, Baltimore MD 21250, USA }

\author{S.B.~Kraemer}
\affil{Institute for Astrophysics and Computational Sciences, Department of Physics, The Catholic University of America, Washington, DC 20064, USA}

\begin{abstract}
The origin of the observed time lags, in nearby active galactic nuclei (AGN), 
between hard and soft X-ray photons is investigated using new {\em XMM-Newton}
data for the narrow-line Seyfert\,I galaxy \object{Ark~564} and existing data for
\object{1H\,0707--495} and \object{NGC~4051}.  
These AGN have highly variable X-ray light curves
that contain frequent, high peaks of emission. 
The averaged light curve of the peaks is directly
measured from the time series, and it is shown that 
(i) peaks occur at the same time, within the measurement uncertainties, at all X-ray 
energies, and 
(ii) there exists a substantial tail of excess emission at hard X-ray
energies, which is delayed with respect to the time of the main peak, and is
particularly prominent in \object{Ark~564}.
Observation (i) rules out that the observed lags are caused by Comptonization 
time delays and disfavors a simple model of propagating
fluctuations on the accretion disk.  Observation (ii)
is consistent with time lags caused by Compton-scattering
reverberation from material a few thousand light-seconds from the primary X-ray source.
The power spectral density and the frequency-dependent phase lags of the peak light curves are consistent with those of the full time series.
There is evidence for non-stationarity in the \object{Ark~564} time series
in both the Fourier and peaks analyses.
A sharp `negative' lag 
(variations at hard photon energies lead soft photon energies) 
observed in \object{Ark~564} appears to be generated
by the shape of the hard-band transfer function and does not arise from
soft-band reflection of X-rays.
These results reinforce the evidence for the existence of X-ray reverberation
in type\,I AGN, which requires that these AGN are 
significantly affected by scattering from circumnuclear material
a few tens or hundreds of gravitational radii in extent.  
\end{abstract}


\section{Introduction \label{introduction}}

Active Galactic Nuclei (AGN) in the X-ray regime can be highly variable over a wide range of timescales. Two common statistical tools for analyzing this variability are the power spectral density (PSD) of the time series, and the lag spectrum, defined as the time lags between light curves obtained from two bands of photon energy and
analyzed as a function of the frequency of Fourier modes \citep[e.g.][]{nowak96a}.
A positive lag is defined as the hard band lagging the soft in a particular frequency band. The lag spectrum provides an important constraint on models that seek to explain the observed variability \citep[e.g.][]{nowak96a, miller10b}

The first clear detection of both positive and negative lags in an AGN was in \object{1H~0707--495} \citep{fabian09a}. At low frequencies ($\nu<5\times10^{-4}$~Hz) the AGN exhibited a positive lag, but at higher frequencies the lag was persistently negative. Similar results have been obtained in other AGN 
\citep[e.g.][]{miller11a, emmanoulopoulos11a, demarco12a}.
The same features are observed in our analysis, presented here, of new observations of 
\object{Ark~564}, which \citet{mchardy07a} have previously suggested displayed 
both positive and negative lags from fits of Lorentzian components to the PSD and cross-spectrum.

A number of models seek to explain the lag spectrum of \object{1H~0707--495}. \citet{fabian09a} and \citet{zoghbi10a} applied the model of \citet{arevalo06b}, invoking propagating fluctuations in the accretion disk to explain the positive lags at low frequencies, together with reverberation of X-rays reflecting from within a gravitational radius\footnote{
Light travels a distance of one gravitational radius ($r_{\rm g} \equiv GM_{\rm BH}/c^2$)
in 10\,s for black hole mass $M_{\rm BH} = 2\times 10^6$\,M$_\odot$.} of a rapidly rotating black hole to explain the high-frequency negative lags. \citet{miller10b} suggested reverberation due to scattering from gas 10s to 100s of gravitational radii from the central black hole, showing how both positive and negative lags may arise from the same process (see also \citealt{miller11a}).  Older work on the positive lags seen in galactic binary systems and AGN also investigated the role of Comptonization up-scattering time delays \citep[e.g.][]{nowak96a, nowak99a}.

Analyses of X-ray AGN time series have usually been restricted to Fourier space, implicitly assuming that AGN time series are statistically weakly stationary and often assuming that they are Gaussian \citep[but see][]{uttley05a}. All the above proposed explanations may be considered as examples of a `moving average' time series \citep[e.g.][]{brockwelldavis}, in which an underlying white noise process has been convolved with a transfer function (considering only the intrinsic source variations free of measurement noise). But the Fourier methods used to date only extract limited information on that transfer function. It is possible to obtain a good estimate of the PSD, and of the lag spectrum between two energy bands, but it is not possible to extract the phase information of the transfer function in the time series for any individual energy band.  This information is vital for reconstructing the transfer functions, and for testing the various models. The aim of this paper is to approach the problem in the time domain, and directly extract the transfer functions from the individual time series by examining the most prominent flares in the light curves. 

In this study, we examine three highly X-ray variable AGN: the nearby
narrow-line Seyfert\,I galaxies \object{Ark~564} (redshift
$z=0.0247$), \object{1H~0707--495} ($z=0.0411$) and \object{NGC~4051}
($z=0.0023$). 
NGC~4051 has a black hole mass $M_{\rm BH} = 1.73^{+.55}_{-.52}\times 10^6$\,M$_\odot$
determined by optical reverberation measurement \citep{denney09a}.
The other two AGN do not have optical reverberation measurements published at the
time of writing.  \citet{botte04a} have estimated the mass of \object{Ark~564}
as $M_{\rm BH}=2.61 \pm 0.26 \times 10^6$\,M$_\odot$ using the estimated
relation between the radius of the broad-line region and luminosity of \citet{kaspi00a}.
\citet{zhou05a} have also used the \citeauthor{kaspi00a} relation to estimate the
mass of \object{1H~0707--495} as $M_{\rm BH} \simeq 2.3\times 10^6$\,M$_\odot$, but
we note that \citet{leighly04a} infers that the mass would be $M_{\rm BH}\simeq 10^7$\,M$_\odot$ 
if \object{1H~0707--495} were radiating at its Eddington luminosity.
The ultraviolet spectra of \object{1H~0707--495} have been extensively modeled
by \citet{leighly04b} and \citet{leighly04a} as being dominated by an outflowing, optically-thick wind
from the accretion disk.
The data used in this paper for \object{1H~0707--495} is
the same time series used by \citet{miller10b}. For \object{NGC~4051}
we use the Suzaku data used by \citet{miller10a}. The \object{Ark~564}
time series is new data from {\em XMM-Newton}.

\begin{table*}
\begin{center}
\caption{Summary of 2011 {\em XMM-Newton} observations of Ark~564.\label{table1}}
\begin{tabular}{llcccccc}
\tableline\tableline
 OBSID         & start date    & duration & exposure & excision & 0.4--1\,keV rate & 4--7.5\,keV rate & 0.4--10\,keV rate\\
               &               & /ks & /ks & $''$ & /s$^{-1}$ & /s$^{-1}$ & /s$^{-1}$\\
 \tableline
0670130201 & 2011 May 24 & 59.5         & 38.6              & 8  & $33.51\pm{0.04}$& $0.752\pm{0.007}$ & $48.79\pm{0.04}$ \\
0670130301 & 2011 May 30 & 55.9         & 34.3              & 0  & $21.67\pm{0.03}$& $0.479\pm{0.004}$ & $31.33\pm{0.03}$\\
0670130401 & 2011 Jun 06 & 63.6         & 31.1              & 0  & $21.56\pm{0.03}$& $0.562\pm{0.004}$& $31.84\pm{0.03}$\\
0670130501 & 2011 Jun 11 & 67.3         & 40.6              & 5  & $26.84\pm{0.03}$& $0.643\pm{0.005}$ & $39.20\pm{0.05}$\\
0670130601 & 2011 Jun 17 & 60.9         & 33.5              & 0  & $22.89\pm{0.03}$& $0.565\pm{0.004}$ & $33.60\pm{0.03}$\\
0670130701 & 2011 Jun 25 & 64.4         & 29.1              & 0  & $14.17\pm{0.02}$& $0.356\pm{0.004}$ & $20.69\pm{0.03}$\\
0670130801 & 2011 Jun 29 & 58.2         & 40.5              & 4  & $23.47\pm{0.03}$& $0.570\pm{0.004}$ & $34.03\pm{0.04}$\\
0670130901 & 2011 Jul 01 & 55.9         & 38.7              & 8  & $30.74\pm{0.04}$& $0.678\pm{0.007}$ & $44.36\pm{0.04}$\\
\tableline
\end{tabular}
\tablecomments{Columns show observation ID, start date, duration of the observation, net exposure time after live time
correction and flaring background removal, excision radius used to correct for pile-up and average count rates after correction for excision, in the 0.4--1\,keV, 4--7.5\,keV, and 0.4--10\,keV bands.}
\end{center}
\end{table*}

\section{Observations and Data Reduction}

\subsection{Ark~564}

For Ark~564 we obtained a set of eight \textit{XMM-Newton} pointed observations, made between May and July 2011. 
The OBSID, the starting date, and the exposure duration of each pointing are reported in the first three columns of Table~\ref{table1}. Given the large energy bandpass and high effective area, we focused our analysis on the European Photon Imaging Camera (EPIC) pn data. All the EPIC pn observations were performed in Small Window mode and used the thin optical blocking filter. 
Data were processed using \textsc{HEAsoft v.6.12} and \textsc{SAS v.11.0.0}.

The raw pn event files were filtered retaining only best-quality (FLAG=0) singles and doubles (PATTERN $\leq 4$) events. 
Source events were extracted from circular regions with a 36$''$ radius, while background events were extracted from two 
boxes of $61'' \times 100''$ and $51'' \times 51''$ size. 
Background light curves were then extracted for single events only, in the $10-12$~keV band, for each observation. 
After an inspection of these light curves, strong flaring background time intervals were removed using a threshold of $0.1$~count~s$^{-1}$. The net exposure times after the flaring background removal is reported in the fourth column of Table~\ref{table1}. These values also account for the Small Window mode live time of $\sim 71 \%$.

The \textsc{SAS} task \textit{epatplot} was then used to estimate the fraction of pile-up affecting the data in the $1-9$~keV band.
We found non-negligible deviations (of the order of $3-5\%$) of the observed pattern distribution from the predicted one in half of the observations. For the affected observations, the cores of the source distribution on the image were excised, using a radius dependent on the amount of pile-up in the data, until
we were able to recover the predicted pattern distribution within a $1-2\%$ uncertainty.  
For the observations where we needed to excise the core (i.e. where we used an annular source extraction region), the source count  was then corrected by a  factor that  represented the point-spread-function correction from the excised annular region to the count rate one would have obtained from a circular source extraction region with no excision. This correction factor ranged from  $0.45-0.75$ and was constant within each individual observation
(Columns 6--8 of Table\,\ref{table1} show count rates corrected by this factor).
Finally, energy dependent exposure maps were generated through the  \textsc{SAS} task \textit{eexpmap}
and the background count rates were corrected for the slightly lower effective exposure found relative to the source region, and also scaled appropriately for each source extraction cell, before the background rates were subtracted from the source light curves. 

\begin{figure}[!h]
\epsscale{1.}
\plotone{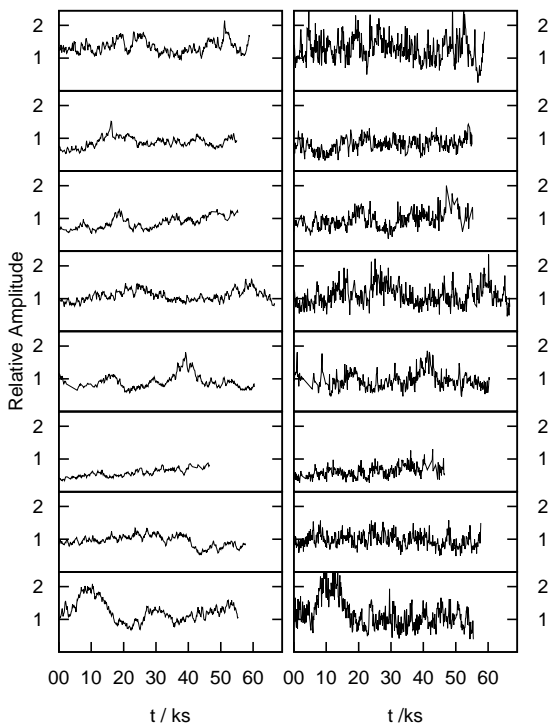}
\caption{Ark~564 light curves in the soft (0.4-1 keV, left) and hard (4-7.5 keV, right) bands for the eight observations (earliest at top). Light curves have been binned to $128$s per time bin, corrected for excision, and scaled relative to the average count rates of 25.7 counts/s and 0.60 counts/s for the soft and hard bands respectively. Scaled 68\% confidence regions are too small to be marked on the plots, but have mean values of 0.026 and 0.17 for the soft and hard bands respectively.\label{fig:lc}}
\end{figure}

Light curves after correction and scaling are shown in Fig.\,\ref{fig:lc}. The count rates have been corrected for the excision performed, and the light curves in each band have been scaled relative to the overall average count rate. The average corrected count rate for the 0.4-1 keV band is $25.7$~s$^{-1}$, and for the 4-7.5 keV band is $0.60$~s$^{-1}$, with typical 68\% confidence regions of 0.026 and 0.17 respectively.

\begin{figure*}[!ht]
\epsscale{2.0}
\plottwo{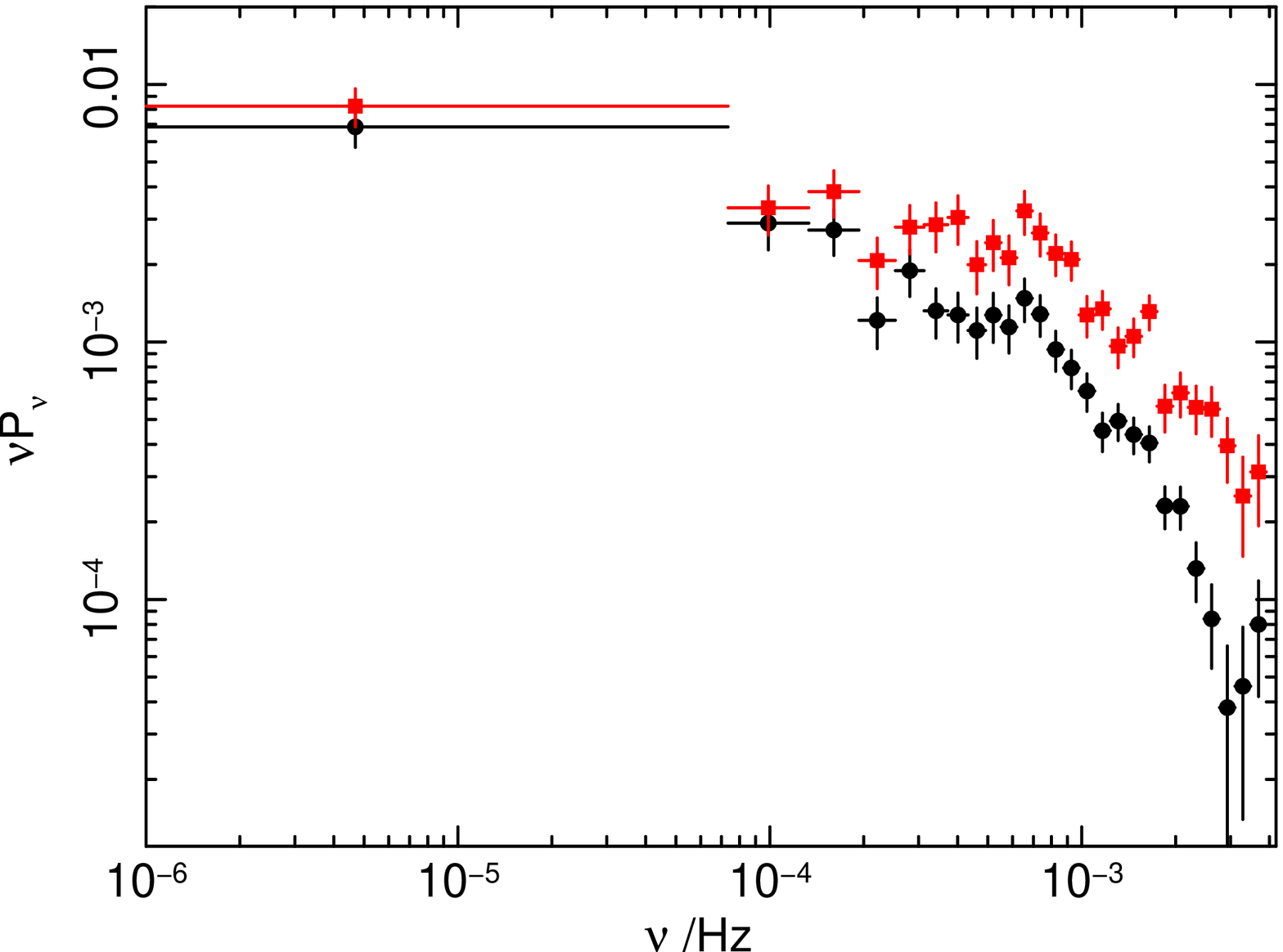}{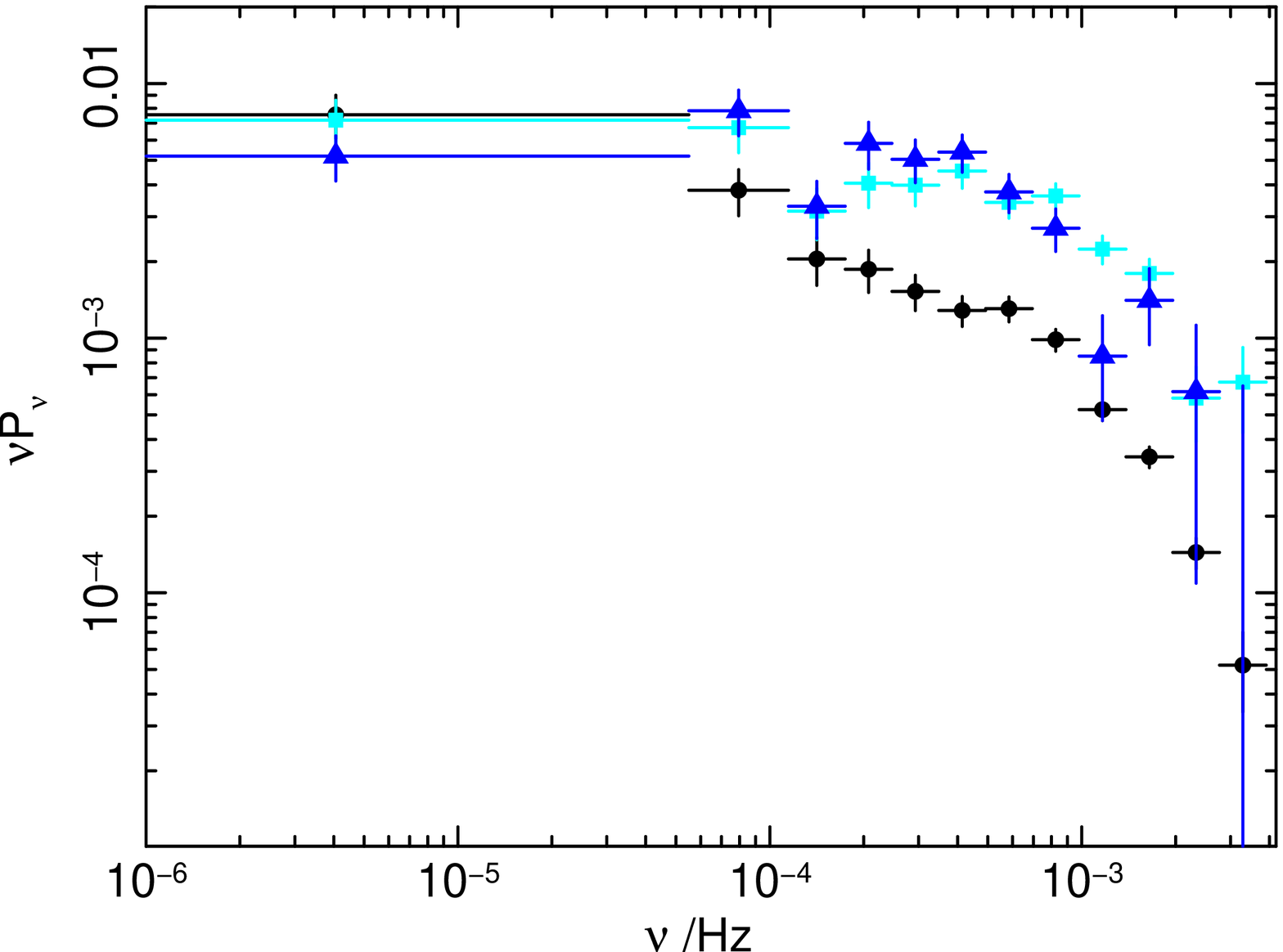}
\caption{
PSDs of Ark~564 
{\em (a) (left panel)} in 0.4--1\,keV (lower circles) 
and 1--2\,keV (upper, squares) bands, sampled at frequency resolution $\Delta\log\nu=0.05$
{\em (b, right panel)} in 0.4--1\,keV (lower circles), 2--4\,keV (squares) and 4--7.5\,keV (triangles) bands,
sampled at frequency resolution $\Delta\log\nu=0.15$.
Vertical bars indicate 68\% confidence region uncertainties,
horizontal bars indicate the range of frequencies incorporated in each point.
\label{fig:psd}
}
\end{figure*}

\subsection{1H 0707-495}

{\it XMM-Newton}  observed  1H\,0707--495 during four epochs, 
2000 Oct, 2002 Oct, 2007 Jan-Feb  and 2008 Jan.  In this paper we present 
analysis of only the largest dataset, from 2008, as in \citet{zoghbi10a}.

In this case, the EPIC observations were made using the medium filter
and with prime large window mode during 2008.  These data products
were based upon the original analysis of \citet{miller10b}, processed
using {\sc SAS v8.0.0} and {\sc HEAsoft v6.8}.  The pattern selections
used were the same as those for Ark~564. The background levels were
high during some of the 1H\,0707-495 observations. See
\citet{miller10b} for details of the filtering of background flares,
and the selection of source and background extraction cells.

\subsection{NGC~4051}

{\it Suzaku} observed NGC~4051 during 2005 Nov 10-13 and 2008 Nov 6-12 and 23-25. Here we analyze only the relatively unobscured 2008 observations \citep[see][]{miller10a}. The {\it Suzaku} X-ray Imaging Spectrometer \citep[XIS,][]{koyama07a} is an ensemble of CCDs that yield useful data over $\sim 0.6 - 10$\,keV, with energy resolution $\sim 150 $\,eV at 6\,keV. Here we use data from the two  front-illuminated units, XIS~0 and XIS~3. 

In brief, the data were reduced using v6.4.1 of HEASOFT.  They were screened to exclude periods during and within 500 seconds of the South Atlantic Anomaly, to  have an Earth elevation angle less than 10$^{\rm o}$  and to have a cut-off rigidity $> 6$\,GeV.  We selected event grades 0,2,3,4 and 6 and cleaned out the hot and flickering pixels. Source products were extracted  from circular regions of $2.9'$ radius with background spectra from a region of the same size, offset from the source. For further detail see \citet{turner09b} and \citet{miller10a}.

\section{PSD and lag spectrum analysis \label{sec:psd}}
In this paper, the time series have been analysed in four bands of photon energy: 
0.4--1\,keV, 1--2\,keV, 2--4\,keV, 4--7.5\,keV, similar to those used in earlier papers.
To test models of reverberation, it would be good to analyze separate line emission
and continuum time series, as in optical reverberation studies.  However, in current
X-ray data, photon statistics are limited and it is not possible to cleanly separate
line emission from continuum: 
it is necessary to work in broad bands of photon energy so that
variations on short timescales are not completely dominated by photon shot noise.

The PSD of \object{1H~0707--495} has previously been shown in \citet{miller10b}.
The PSD of \object{NGC~4051} for this dataset has been shown by \citet{miller10a}
as well as in previous work by \citet{mchardy04a}.  \object{Ark~564} has previously
been studied by \citet{arevalo06a} and \citet{mchardy07a}. 
Fig.\,\ref{fig:psd} shows the PSD of the new observations of
\object{Ark~564} in each band, obtained using the
maximum-likelihood method of \citet{miller10b, miller10a} which
corrects for the gaps in the time series caused by breaks between
observations and the filtering out of bad data during observations,
and also for the presence of shot noise. The PSDs are normalized to the mean of the full series. 
The method provides accurate estimates of the PSD uncertainties and of the covariance between
frequency channels, although the latter are difficult to convey in
simple diagrams. The two softest bands have high count rates and their
PSDs are shown at high frequency resolution in Fig.\,\ref{fig:psd}(a).
At high frequencies the frequency channels are largely independent and
have widths $\Delta\log_{10}\nu=0.05$.  The maximum frequency shown is the Nyquist
frequency corresponding to the chosen time series sampling of 128\,s.  At low frequencies
the finite duration of the individual observations causes nearby Fourier modes to become
correlated, so the frequency bin widths have been increased to preserve approximately
the statistical independence of the plotted points.  The PSDs of the two harder bands
are shown at lower frequency resolution in Fig.\,\ref{fig:psd}(b), because of the larger
statistical uncertainties. We also show the $0.4-1$\,keV band at this resolution, for reference. 

We note in the Ark~564 PSDs the 
features previously reported by \citeauthor{arevalo06a, mchardy07a}, namely the very steep
fall-off to high frequency and the complex shape of the PSD, which appears not be described
by a simple smoothly-varying quasi-power-law form.  There appears a sharp feature at
frequency $\nu \simeq 7 \times 10^{-4}$\,Hz which is too ill-defined to be claimed as a
`quasi-periodic oscillation', but at least does indicate the presence of a sharp transition in the PSD.
The significance of the potential feature depends on the model fit to the PSD. A smoothly-varying quasi-power-law would imply a significant excess of power, whereas fitting a broken power-law would result in no apparent excess in power.
We do not explicitly consider the origin of this feature in this paper. 
As previously found, and as commonly observed in other AGN, the PSD is higher for the time series of higher energy 
bands at high frequencies, although they are more comparable in amplitude at low frequencies.

\begin{figure}[!h]
\epsscale{1.0}
\plotone{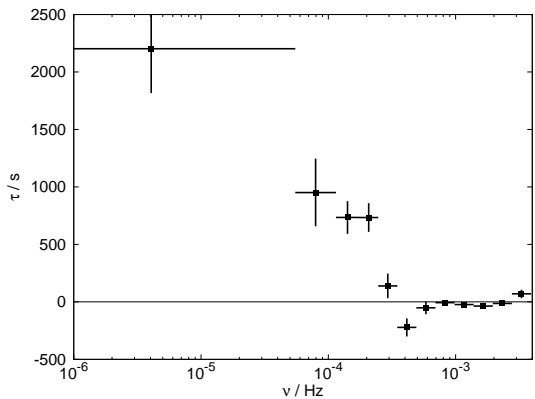}
\caption{Lag spectrum of Ark~564 between the 0.4--1\,keV and 4--7.5\,keV bands at frequency
resolution $\Delta\log_{10}\nu=0.15$. Vertical bars indicate 68\% confidence region uncertainties,
horizontal bars indicate the range of frequencies incorporated in each point.
\label{fig:lagspectrum} }
\end{figure}

Lag spectra between various bands have been shown for \object{1H~0707--495} and {NGC~4051} by
\citet{fabian09a, zoghbi10a, miller10b, miller10a} and for previous observations of \object{Ark~564} by
\citet{arevalo06a, mchardy07a, demarco12a}.  The lag spectrum between the 0.4--1\,keV and 4--7.5\,keV bands
from the new observation of \object{Ark~564} is shown in Fig.\,\ref{fig:lagspectrum} with
frequency resolution $\Delta\log_{10}\nu=0.15$.  There are strong positive lags up to $\tau \simeq 2000$\,s
at low frequencies, as previously reported by \citet{arevalo06a}, and negative lags at higher frequencies.  
We note the sharp negative lag of $\tau = -222 \pm 78$\,s 
at $\nu \simeq 4 \times 10^{-4}$\,Hz accompanied by oscillatory and negative lags, $\tau \simeq -20$\,s,
at higher frequencies, similar to the simple top-hat reverberation transfer functions discussed
by \citet{miller10b} and \citet{miller11a}. Overall, the lag spectrum has a closely similar behavior to that
previously seen in well-studied AGN such as \object{1H~0707--5495} \citep{miller10b}, and that similarity
suggests that there is a single physical mechanism common to all these AGN, a point we return to in
section\,\ref{sec:discussion}.

The frequency of the sharp negative lag and the previously cited mass of Ark~564 are in line with the frequency-mass relation of \citet{demarco12a}. The magnitude of the lag is much greater than the \cite{demarco12a} lag-mass relation would imply. However, the hard band used by \cite{demarco12a} is 1-5~keV, very different from the 4-7.5~keV hard band used in this analysis.

\section{Non-stationarity of the PSD and lag spectra}\label{sec:nonstationary}
\begin{figure*}[!ht]
\epsscale{2.0}
\plottwo{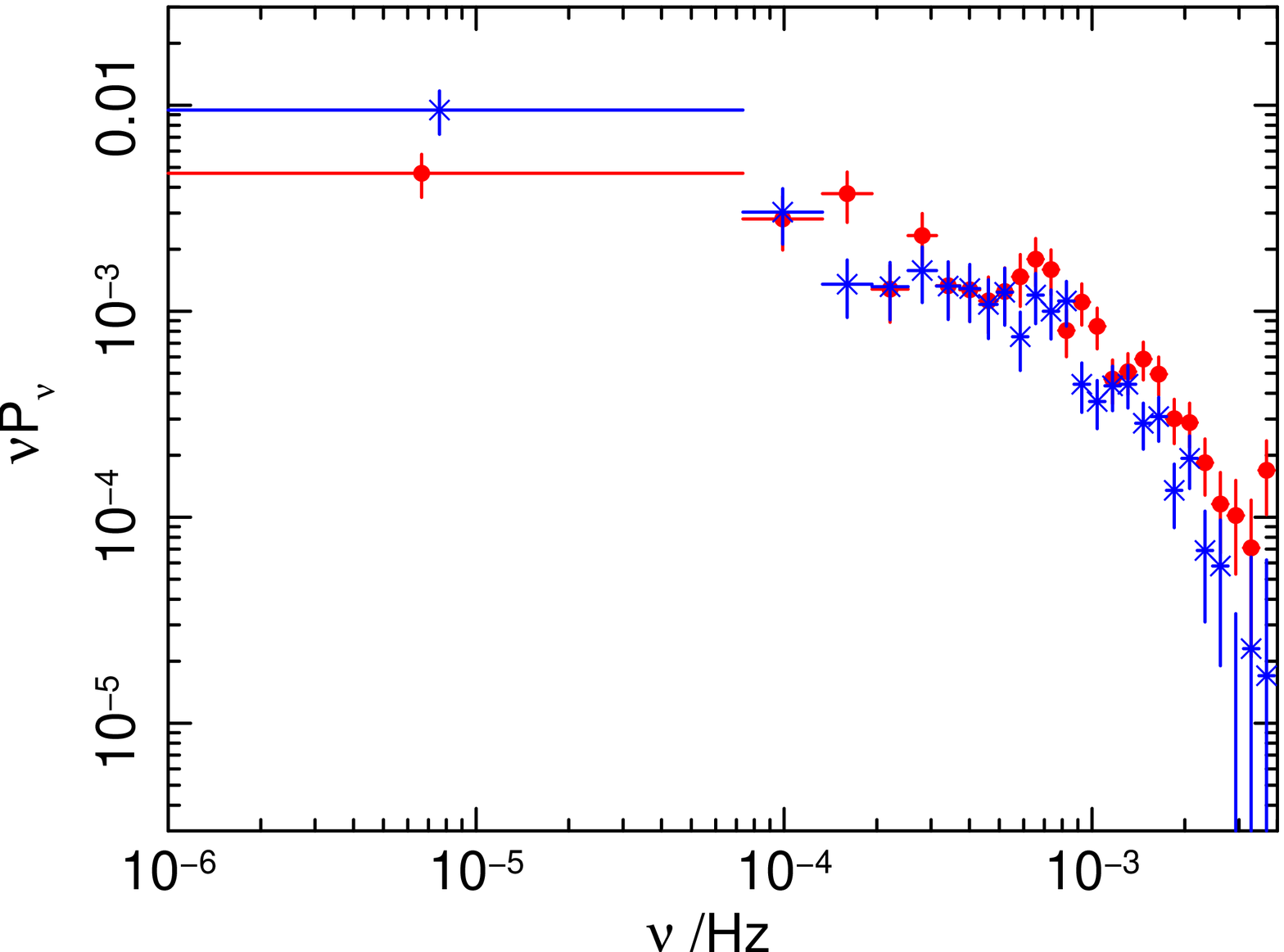}{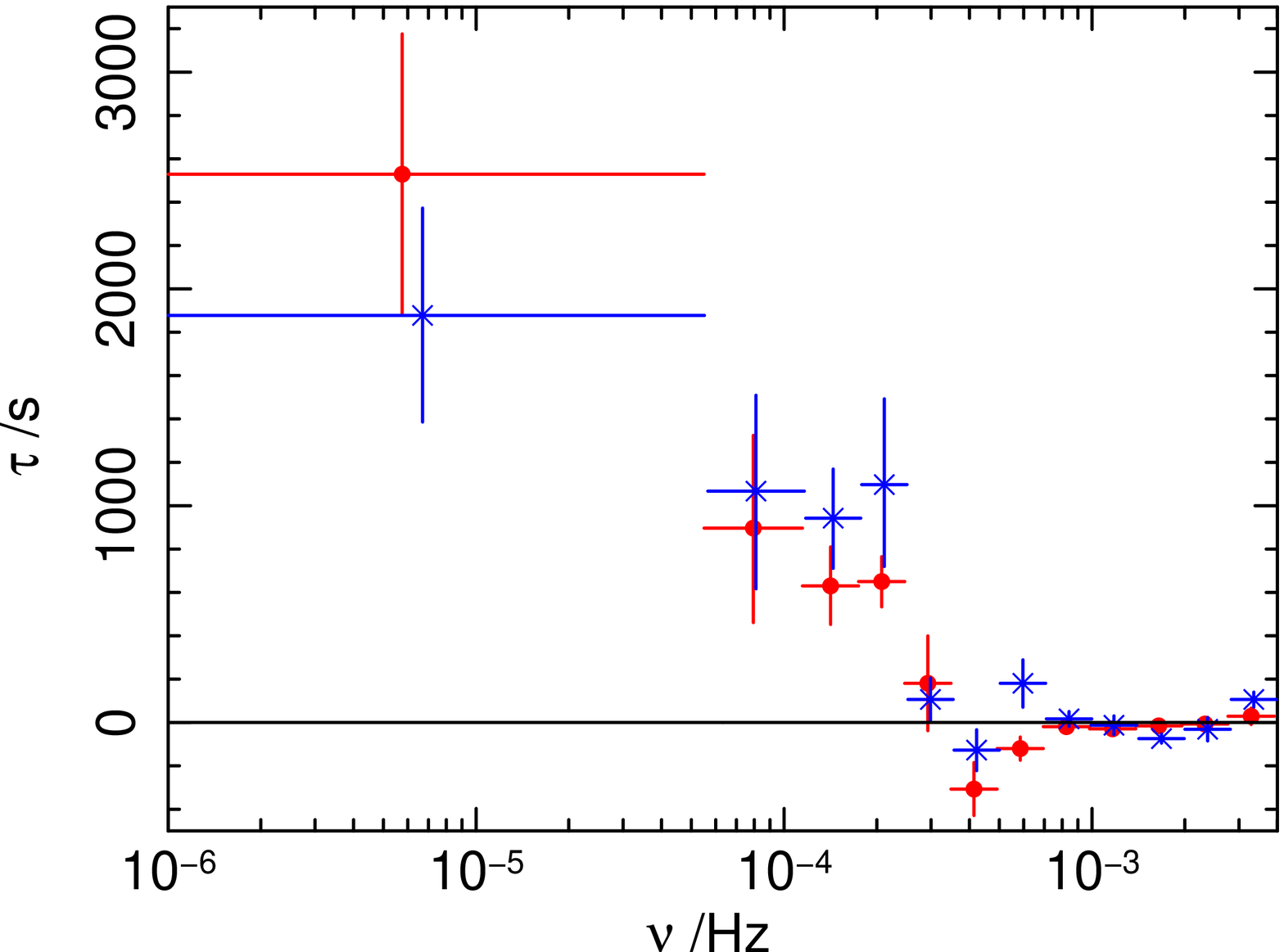}
%\resizebox{0.48\textwidth}{!}{
%\rotatebox{0}{
%\includegraphics{}
%}}
%\resizebox{0.48\textwidth}{!}{
%\rotatebox{0}{
%\includegraphics{}
%}}
\caption{
{\em (left)} PSD of Ark~564 in the 0.4--1\,keV band, sampled at frequency resolution $\Delta\log\nu=0.05$,
analyzed for the first four OBSIDs (filled circles) and the last four OBSIDs (crosses).
{\em (right)} Lag spectra between the 0.4--1\,keV and 4--7.5\,keV bands,
sampled at frequency resolution $\Delta\log\nu=0.15$, for the same data subsets.
Vertical bars indicate 68\% confidence region uncertainties,
horizontal bars indicate the range of frequencies incorporated in each point.
\label{fig:nonstationary}
}
\end{figure*}

In the course of the analysis of the light curve peaks described in the following sections,
some evidence was noticed for non-stationarity of the time series (Section \ref{sec:shape-ark-byhalf}).  In this section we test for non-stationarity in the powerspectrum analysis by dividing the set of observations
into two halves and recalculating the PSD and lag spectra.  Fig.\,\ref{fig:nonstationary}
shows the PSD for the soft band $0.4-1$\,keV only, and the lag spectra between the
soft and hard, $4-7.5$\,keV, bands, for the first four and last four OBSIDs analyzed independently.
There is clear evidence for the PSD having changed in
amplitude on a timescale of a few weeks between these two sets of observations (see Table\,\ref{table1}),
and some evidence for the shape having changed. 

The lag spectra also appear to have varied: the sharp negative feature
appears more pronounced with a lag $\tau = -306\pm 122$\,s in the first half but less pronounced
in the later half, $\tau = -127\pm 94$\,s: this difference on its own is not very significant but
is worth noting, given the other apparent departures from stationarity that are seen.
The negative lags at high frequency in the overall lag spectrum
(Fig.\,\ref{fig:lagspectrum}) appear not to be present in the earlier half and to be more
oscillatory in the later half.  We also analyzed each of the eight OBSIDs independently, but given
the larger uncertainties did not find significant evidence for non-stationarity on these shorter
timescales.

Apparent variations in estimated PSDs may in principle arise from red noise leak of power below the lowest sampled frequency, causing what may be termed an apparent `weak non-stationarity' \citep[e.g.][]{vaughan03a}. For the observations presented here, the lowest frequencies measured are more than two decades lower than the PSD break frequency, and the time series variance only weakly diverges at the low frequency limit.  The PSD variations that are observed occur throughout the full frequency range, and do not preferentially occur at low frequencies. It is likely that they reflect genuine non-stationarity, rather than arising from the presence of inadequately-sampled long-timescale variations.  This conclusion is reinforced by the changes observed in the time-domain transfer function (section \ref{sec:shape-ark-byhalf}).

\section{Analysis of light curve peaks}

\subsection{The goal of the peaks analysis}
While a number of claims have been made for observation of X-ray
reverberation in AGN, either with a measured time delay \citep[e.g.][]{fabian09a, miller10a}
or without \citep{miller06a, ponti12a},
no studies have extracted the associated transfer
function in an individual energy band. For those claims with a measured time delay,
the lag spectrum has been derived from the 
phases of the cross-spectrum, which itself is the Fourier transform of the 
cross-correlation function.  Previous studies have measured only the relative
phases of modes between two bands, which does not allow a unique interpretation
of the measured relative time lags.  The three mechanisms for time delays proposed
to date (Comptonization delays, reverberation, and fluctuations propagating radially across 
an accretion disk)
are all expected to be characterized by an energy-dependent transfer function.
To measure the transfer function in an individual time series requires a
blind deconvolution algorithm. Extensive work has been done on the
problem of blind deconvolution, mostly for image analysis (see
\citealt{kundur96a} for an overview). However, the presence of significant shot noise and gaps
in the time series rules out the use of many of the simpler techniques. 

Instead of performing a full deconvolution, we look at the behavior
of the time series in the region of flares in activity. The presence
of a flare brings the emission significantly above the level of the
complicated `long-timescale' temporal activity. By examining the temporal response 
to such a flare in each energy band, we can obtain an estimate of the shape of
the transfer function. Individual flares suffer too much noise, arising from the source variations, to provide useful information on the transfer function, but by averaging many flares we can obtain an estimate, as shown below. This is similar to the technique of `flare stacking' briefly described, but not significantly used, by \citet{zoghbi10a}.

\subsection{The selection and analysis of light-curve flares\label{sec:peakselection}}
Flares in the time series were selected for the analysis and measured by a
multi-stage process. We considered a model in which the light curve
comprises a long-timescale `baseline' variation, which we wish to filter out, 
on which are superimposed 
flares, with some unknown temporal structure, which we wish to measure.  
In practice, the long-timescale variation itself is likely to be composed of many overlapping
flares, but we are unable to separate those into individual events. The more extreme flares 
are brought above the baseline level, and we hope to use this to examine their shape.  Peaks in the light curve were
first selected by median filtering the time series, as follows.

\newcounter{counter1}
\begin{list}{\arabic{counter1}.~}{ \usecounter{counter1}
		 \setlength{\itemsep}{1pt}
     \setlength{\parsep}{1pt}
     \setlength{\topsep}{0ex}
     \setlength{\partopsep}{0pt}
     \setlength{\leftmargin}{1.em}
     \setlength{\labelwidth}{2.em}
     \setlength{\labelsep}{0.em} }
\item The raw soft-band time series was filtered with a median filter of length $2R+1$ time bins, 
giving a first-level baseline and residual time series. %1

\item The first-level residual time series was smoothed with a smoothing function derived from the soft-band PSD. %2

\item Peaks of greater than $2\sigma$ significance were identified in the smoothed time series. %3

\end{list}

A median filter is better than a linear 
filter at preserving temporal structure in the residual time series on timescales
less than half the length of the filter, but to accurately recover that structure,
it is necessary that the
length of the filter window be greater than the size of any
features under investigation.  The lengths of the filter for this detection
stage are given in Table\,\ref{tbl:params}, and are significantly longer than the 
FWHM of the detection smoothing function, as required.

The choice of smoothing function affects which peaks are selected,
with a tendency to select peaks whose shape and duration are similar to
those of the smoothing function.  To optimize the choice of smoothing
function, we proceeded as follows.  In the `moving average' model of
the time series, if the underlying time series has a white noise powerspectrum,
the observed PSD is the modulus-squared of the
Fourier transform of the transfer function.  Assuming that the
transfer function is symmetric in the 0.4--1\,keV band,
we obtained an estimate of the shape and
length of the transfer function by Fourier-transforming the positive,
real square-root of the PSD.  The assumption of symmetry is important
as we should avoid introducing spurious time lags into the analysis
through our choice of smoothing function.
Because the peaks were detected in the
residual time series after the data had been median filtered, we also
applied the same procedure to the PSD-derived function to obtain the
final smoothing function.  The FWHM of the resulting functions were
$500, 530, 370$\,s for each of \object{NGC~4051},
\object{1H~0707--495} and \object{Ark~564}, respectively. 
While the derived FWHM are rather similar between these AGN, 
this is a reflection of the similarity of the shapes of their PSDs, 
and in particular of their break frequencies.

To test the sensitivity of the results to the choice of smoothing function,
we also experimented 
with a Gaussian smoothing function with the same FWHM. We obtained
similar results, although the detailed shapes of the
primary flare peaks differed slightly in the two cases, and the
powerspectra of the selected peaks (section\,\ref{sec:peakspectralanalysis}) 
exhibited small differences.

Having selected a set of peaks in the soft band, we then attempted to
characterize their temporal shapes in the soft and hard bands. To sample the
long, ks, timescales present in the lag spectra, a long filter would be required.
However, directly applying a long median filter to the raw data
produced unsatisfactory baseline curves, which did not follow the 
light curve on long timescales owing to the substantial structure on 
intermediate timescales. The solution adopted was nested
filtering, in effect performing a form of wavelet median filtering. 
The measurement steps continued from the peak-detection stage, as follows.

\begin{list}{\arabic{counter1}.~}{ \usecounter{counter1}
		 \setlength{\itemsep}{1pt}
     \setlength{\parsep}{1pt}
     \setlength{\topsep}{0ex}
     \setlength{\partopsep}{0pt}
     \setlength{\leftmargin}{1.em}
     \setlength{\labelwidth}{2.em}
     \setlength{\labelsep}{0.em} }
\setcounter{counter1}{3}
\item The first-level baseline from the peak-detection stage was median filtered with length $4R+1$ time bins. %4

\item The result was further median filtered with length $8R+1$ time bins, giving a third-level baseline. %5

\item This third-level baseline was subtracted from the original time series to give a third-level residual time series. The three-stage filtering process produced baselines that better fit the behavior of the light curve than a single stage filter of length $8R+1$ time bins.%6

\item For each of the peaks detected in step 3, a region $8R+1$ time bins in width, centred on each peak, 
was selected from the third-stage residual, to measure its shape. %7

\item A weighted average of all peak shapes was performed to give an estimate of the average peak shape. %8

\item Steps 1, 4-8 were performed on the hard-band time series to give an estimate of the average hard-band peak shape,
using, in step 7, the same peak times that were found for the soft band. %9
\end{list}

\begin{figure*}
\epsscale{1.8}
\plotone{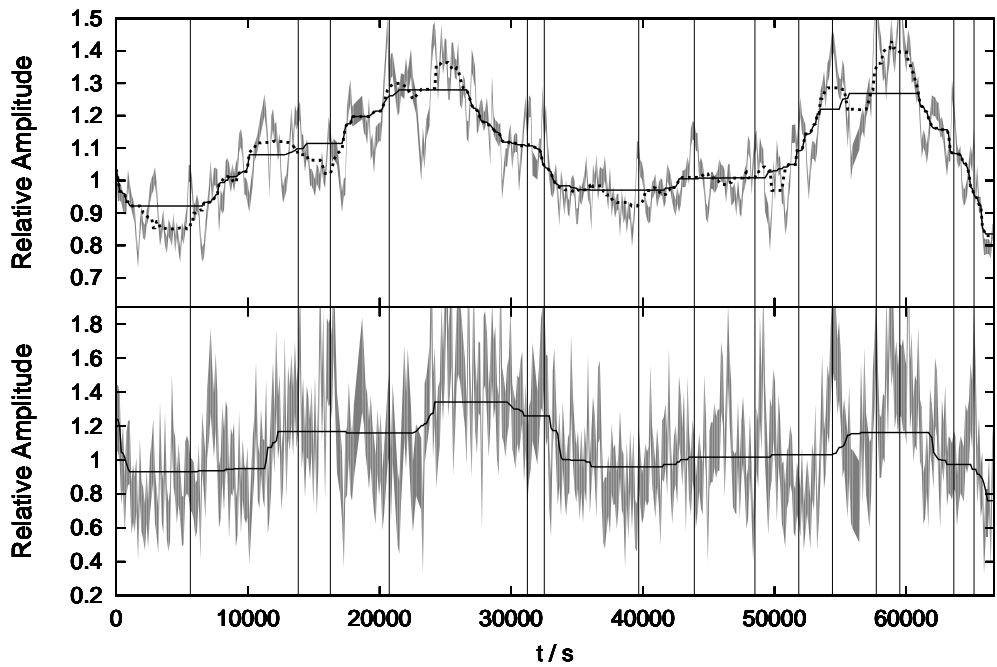}
\caption{A section from the Ark~564 time series. The upper panel shows the soft (0.4--1\,keV) band, while the lower panel shows the hard (4--7.5\,keV) band. Values have been scaled by the overall mean flux in each band, and the 68\% confidence region due to shot noise is indicated by the grey shading. The solid black curve denotes the third-stage median filter baseline. The black dotted line in the upper plot shows the first stage median filter baseline used for peak selection. 
Vertical lines show the times of peaks selected in the soft band.
\label{fig-light-curve}}
\end{figure*}

Because we are interested in the mean shape of peaks of differing amplitude, 
each individual peak and its neighboring filtered values needed to be
scaled. This was achieved by dividing each set of values, in the soft and hard
bands, by the characteristic height of the soft-band peak. The
characteristic height was found by smoothing the final residual time series with
the same smoothing function used to select the peaks, to reduce the impact of shot noise on the central peak
value, and measuring the maximum value.
The final average shape in each band was then calculated using a weighted average 
of the individual, scaled but unsmoothed, peak shapes. The weights used were the total shot-noise variance in the soft band.

The result is that the hard band peak shape is generated using the peak times, scalings and weights of the soft band peaks, 
and is completely blind to the hard band light curve.

\subsection{Peak Shapes}

\begin{figure*}
\epsscale{2.5}
\plotone{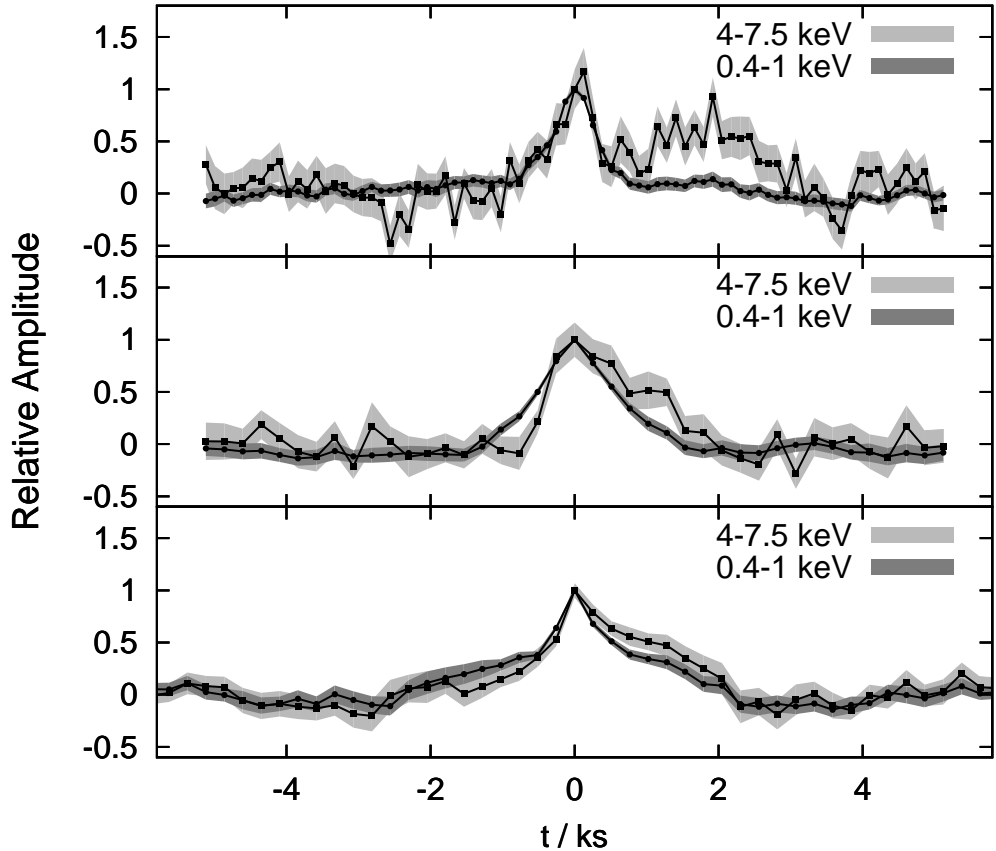}
\caption{The extracted peak shapes for the three AGN: 
(top) Ark~564 sampled at 128\,s; 
(centre) 1H~0707--495 sampled at 256\,s; 
(bottom) NGC~4051 sampled at 256\,s.
Shaded regions indicate the 68\,percent confidence region for each plotted point.
Shapes are shown for the 0.4--1\,keV (dark shading) and 4--7.5\,keV (light shading) bands. 
The peak shapes have been normalized so that the amplitude at $t=0$ is $1.0$. \label{fig-shapes}}
\end{figure*}

The resulting peak shapes are shown in Fig.\,\ref{fig-shapes}. The
peak shapes in each band have been scaled such that the height at
$t=0$ is unity. This allows for easier visual comparison of the
shapes.  In \object{Ark~564}, an approximate comparison of the hard
band primary peak with the excess gives a unitless area for the
primary of $4.6\pm 0.4$ over a range in time of $\pm 384$\,s, while
the excess has an area of $7.6 \pm 0.7$ over a range in delay time of
640--2700\,s (the appropriate time range was obtained from visual
inspection of Fig.\,\ref{fig-shapes} and is rather uncertain).  Thus
the hard band excess contains about $1.65 \pm 0.2$ times as many
photons as the hard band primary peak. In \object{NGC~4051} and
\object{1H~0707--495} the delayed excess and the primary peak cannot
be easily separated, making it difficult to evaluate the fraction of
delayed photons. Visual inspection indicates that those fractions are
smaller in these two AGN compared with \object{Ark~564}.  These values
and the shapes obtained are discussed in
section\,\ref{sec:discussion}.

\begin{table*}
\begin{center}
\caption{AGN peak detection and measurement parameters\label{tbl:params}}
\begin{tabular}{|l|rrr|}
\tableline
AGN & \object{Ark~564} & \object{1H~0707--495} & \object{NGC~4051} \\
\tableline
Sampling time /s & 128 & 256 & 256 \\
Peak detection filter length /s   & 2688    & 2816 & 5376 \\
Peak measurement filter length /s & 10368   & 10496 & 20736 \\ 
Smoothing function FWHM /s	 &	370	&	530	&	500	\\
Peak Significance	       & $2\sigma$	&	$2\sigma$	&	$2\sigma$	\\
\tableline
Number of peaks found		&	$117$	&	$65$	&	$79$	\\
\tableline
\end{tabular}
\end{center}
\end{table*}

The parameters used to obtain the peak shapes are shown in Table\,\ref{tbl:params}.  
The shapes of the peaks were relatively insensitive to changes in the parameters. The time series for AGN \object{1H~0707--495} and \object{NGC~4051} were binned in intervals of $256$\,s owing to the noisiness of the data. \object{Ark~564} has better quality data, and was binned at $128$\,s. Reducing the size of the time bins results in noisier peak shapes, and difficulties in obtaining clear PSD and lag spectra (section\,\ref{sec:peakspectralanalysis}). Increasing the bin size reduces the resolution available on the peak shape. The bin sizes were chosen to provide a good compromise between these two factors. The values chosen oversample the peak by a factor 2-3, while the delayed excess is on a much longer timescale, and thus well sampled.

Since we are only sensitive to timescales shorter than the filter window length, 
the first-stage filter length was chosen to be sufficiently long that
modes at frequencies below the break in the PSD were included.  The third-stage filter length
was a factor four larger, such that the low frequency modes that show positive lags in the
lag spectrum were included.  In the case of \object{NGC~4051} it was found that the filter
lengths chosen resulted in significant negative residuals appearing in the final mean peak
shape, which disappeared when the filter lengths were increased by a factor 2, so in this case
that longer filter length was adopted.

The main other effect of increasing the filter length is to increase the significance of the hard-band excess, and to increase the number of peaks detected. The latter effect is because as the filter length increases, the variations about the baseline level increase and more peaks meet the significance criteria.  The former effect likely arises because more low-frequency, positive-lag modes are included as the filter length is increased.

As we were concerned about the possible impact of overlapping neighboring peaks, we performed a visual inspection of the peaks located. While some peaks do overlap, the overall impact is symmetric - for every peak with another closely following it, there is a peak with one closely preceding. The final impact of the few overlapping peaks is minimal.

\subsection{Peak PSD analysis \label{sec:peakspectralanalysis}}

We would like to verify that the peak shapes we extract are a good estimate of the transfer function. To do this, we Fourier transform the peak shapes shown in Fig.\,\ref{fig-shapes}, deduce their power and lag spectra and compare those with those from the full time series. If the time series is a simple, stationary convolution of an underlying process with a transfer function, then we expect the spectra to have similar features.

To find power and lag spectra, the peak shapes were padded to a length
of $1001$ time bins (i.e. much longer than the timescale of interest
to avoid aliasing problems), and Fourier transformed. The Fourier
modes were then binned to equal sized bins in log-frequency. The
bin widths were sufficiently wide to eliminate correlation between adjacent
modes. The low frequency information created in the padded series below the lowest measured values
was discarded.  The padding was important for eliminating aliasing caused by the periodic
boundary condition of the Fourier transform, and prevents spurious modes being created within
the frequency range of interest.

\begin{figure*}
\epsscale{2.5}
\plotone{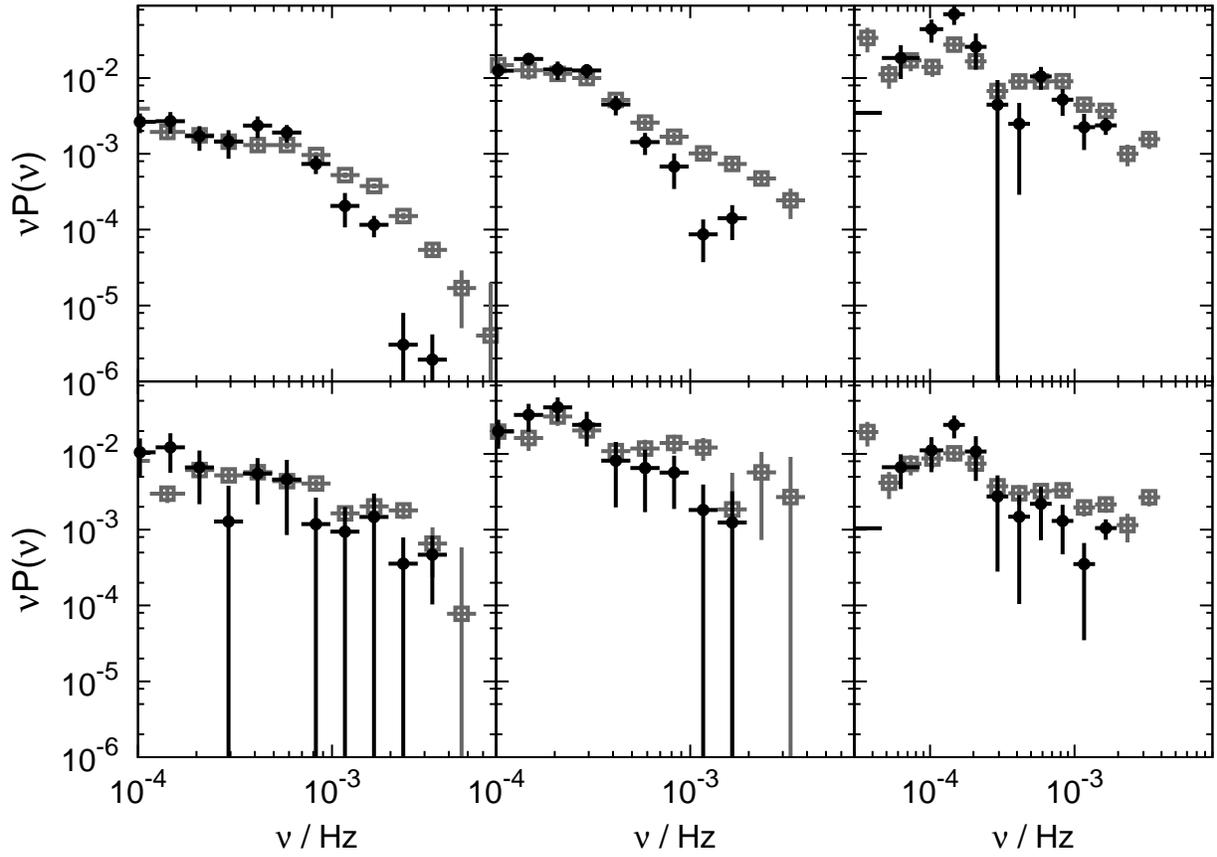}
\caption{Power spectra from the extracted peak shapes of the three AGN: (left) Ark~564; (center) 1H~0707--495; (right) NGC~4051. The top three panels display the soft (0.4--1\,keV) band power spectra, while the bottom three panels show the hard (4--7.5\,keV) band power spectra. Each panel compares the power spectrum of the peak shape  we extract (black circles) with the PSD of the full time series (gray squares). Vertical bars indicate the 68\% confidence region on each point.
\label{fig-ps}}
\end{figure*}

Bootstrap estimates were made of the statistical uncertainties on the Fourier modes. For
an averaged peak shape formed from a weighted average of set of $n$ individual peaks, the bootstrap method involved generating 1000 realizations, each consisting of $n$ peaks drawn at random, but with replacement, from the observed set. In a given realization, some peaks could be present multiple times, others not at all. For each realization, peak shapes and Fourier transforms were found and
the variance of the realizations was taken to be the variance on the Fourier transform of the actual peak shape. The resulting PSDs are shown in Fig.\,\ref{fig-ps}, compared with the PSDs of the
raw time series for the same observations.  Because the peaks transfer function
has been been renormalized, and in any case by selecting the highest peaks the amplitude cannot
be considered typical of the entire time series, the transfer function PSDs have been normalized
so that the total variance matches that of the full time series variance. It can be seen that the
PSD of the transfer function and of the full time series match well for \object{NGC~4051}, and also
match well at low frequencies in \object{1H~0707--495} and \object{Ark~564}, but that the transfer
function has less power at high frequencies in the latter AGN, particularly in the soft band.
This is discussed further in section\,\ref{discussion:psd}.

\begin{figure*}
\epsscale{2.5}
\plotone{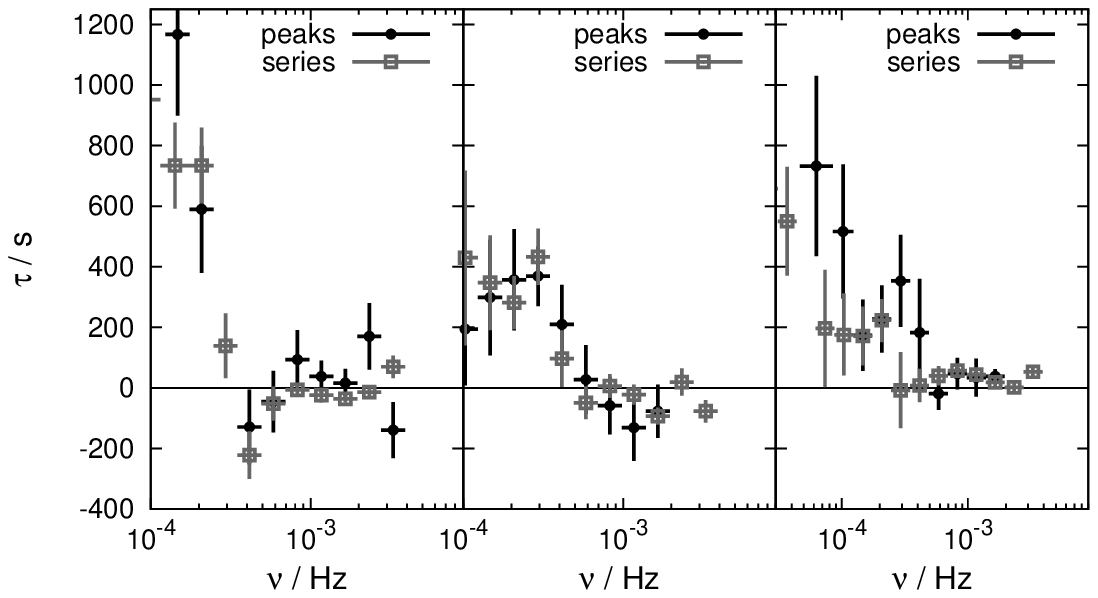}
\caption{Lag spectra between $0.4-1$\,keV and $4-7.5$\,keV bands from the extracted peak shapes (black points) compared with that derived from the full time series (gray points) for the three AGN: (left) Ark~564; (center) 1H~0707--495; (right) NGC~4051. Vertical bars indicate the 68\% confidence region on each point.
\label{fig-lags}
}
\end{figure*}

Lag spectra between the hard ($4-7.5$\,keV) and soft ($0.4-1$\,keV) bands were derived from the transfer functions, and are compared with lag spectra from the full time series in Fig.\,\ref{fig-lags}. 
Uncertainties were estimated using the bootstrap process described above. The transfer function lag spectra are more noisy, 
as expected given that only a subset of the data has been used, but reproduce well the principal
features at low frequencies, including the transition to lags around zero at high frequency.
With the exception of the sharp negative lag feature in \object{Ark~564}, the transfer
functions are too noisy to be able to reproduce the negative lags at high frequencies.

Overall, we conclude from the similarity of both the PSDs and the lag spectra 
of the extracted peaks and the full time series demonstrate that the peaks analysis
method does capture both the powerspectrum and lag information that is obtained from a
standard Fourier analysis, but now with the benefit of being represented as the time-domain
transfer function in an individual energy band.

\subsection{The dependence on energy band of peak selection \label{sec:waveband}}
\begin{figure*}
\epsscale{2.5}
\plotone{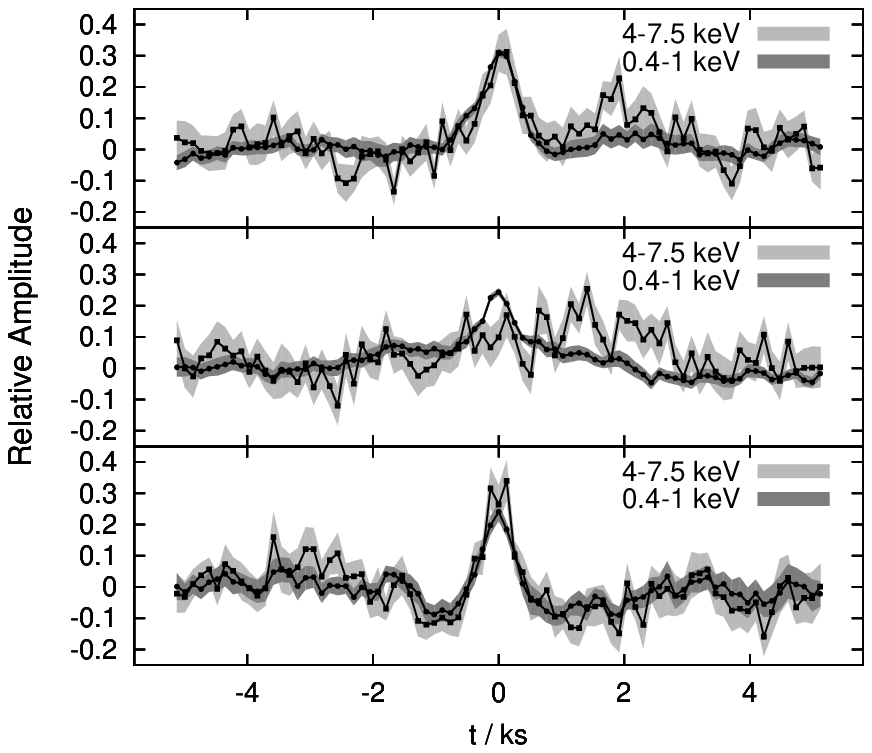}
\caption{Peak shapes in soft and hard bands for three different selections of peaks in Ark~564: (top) peaks detected in both soft (0.4--1\,keV) and medium (1--2\,keV) bands; (center) peaks detected in only the soft band; (lower) peaks detected in only the medium band. 
Shaded regions indicate the 68\,percent confidence region for each plotted point.
Shapes are shown for the 0.4--1\,keV (dark shading) and 4--7.5\,keV (light shading) bands. 
\label{fig-oddity}}
\end{figure*}

Considering the extreme appearance of the hard excess in \object{Ark~564},
we undertook a further analysis to examine its nature. To do this we
also investigated time series in a medium energy band (1--2\,keV). When we perform the peak
selection in this medium band, the resulting set of peaks is different
from those selected in the soft band.  We end up with three sets of
peaks - 52 detected only in the soft band, 29 detected only in
the medium band, and 65 detected in both. These peak shapes are shown in Fig.\,\ref{fig-oddity}.
The hard-band peak shape for this latter category shows a hard excess similar to that found previously. 
The set of peaks found only in the soft band produces an
extreme hard-excess, with a small primary peak. The set of peaks found
only in the medium-band shows no sign of a hard excess. A close
examination of the peak positions shows that while no peak was
detected in the soft-band, the positions are those of maxima in the
soft-band, just not significant enough to make the cut for selection. Whether this is due to noise or a systematic weakening of a subset of peaks in the soft-band is unknown.
This analysis is discussed in section\,\ref{sec:xray_reverb}.

\subsection{The effects of non-stationarity on the peak shape}\label{sec:shape-ark-byhalf}

\begin{figure*}
\epsscale{2.}
\plotone{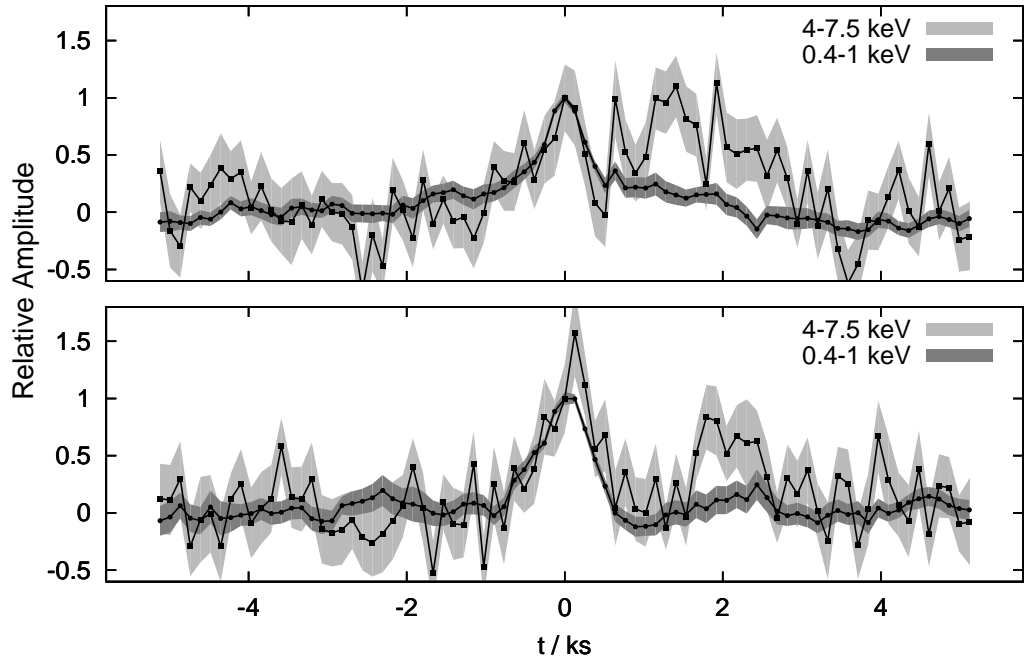}
\caption{Peak shapes in soft and hard bands for the first four observations (top) 
and the last four observations (bottom) of Ark~564. 
Shaded regions indicate the 68\,percent confidence region for each plotted point.
Shapes are shown for the 0.4--1\,keV (dark shading) and 4--7.5\,keV (light shading) bands. 
\label{fig-shapes-ark-byhalf}}
\end{figure*}

In section\,\ref{sec:nonstationary} we have shown that the underlying
process in \object{Ark~564} is non-stationary. In light of this we
examine peak shapes obtained from each of the two halves of the
dataset. The resulting peak shapes, shown in
Fig.\,\ref{fig-shapes-ark-byhalf}, exhibit clear differences. Both
peak shapes exhibit an excess in the hard band at later times than the
primary peak. The shape from the first half of the data set has a
larger excess, peaking with an intensity comparable with that of the
primary peak and extending over $\sim 2$ks. A crude comparison of the
hard band primary peak with the excess gives a unitless area for the
primary of $3.9\pm 0.7$ over a range in time of $\pm 384$\,s, 
while the excess has an area of $10.9 \pm 1.1$
over a range in delay time of 900--3000\,s. This corresponds to the
hard band excess containing about $2.8 \pm 0.6$ times as many photons
as the hard band primary peak.  The shape from the second half of the
data set has a far smaller hard excess extending over just $\sim 1$ks.
The approximate comparison of the hard band primary peak with the
excess gives a unitless area for the primary of $5.5\pm 0.6$ while the
excess has an area of $3.5 \pm 0.7$ over a range in time of 1500--2700\,s, 
corresponding to the hard band
excess containing approximately $0.6 \pm 0.15$ as many photons as the
primary peak (the time ranges were chosen following visual evaluation
of the period having the most significant delayed excess).

It is unclear how this effect and that shown in section
\ref{sec:waveband} are linked. Table \ref{tbl:peaks} lists the number
of peaks detected in each of the observation periods. In the first
four observations, peaks detected only in the soft ($0.4-1$\,keV) band outnumber those
detected in both soft and medium ($1-2$\,keV) bands. In the last four observations
the converse is true. Peaks detected only in the medium band appear to be
uniformly spread.  Thus the primary cause of
the apparent variations in the delayed excess is not clear. Does the excess depend primarily
on the peak spectral hardness ratio, with peaks of differing hardness ratios
appearing in the time series preferentially at differing epochs? Or are both
the variations in the delayed excess and the variations in the hardness of the detected
peaks separate manifestations of the non-stationary variations in the source?

\begin{table*}
\begin{center}
\caption{The number of peaks found in each OBSID, broken down by detection only in the soft ($0.4-1$\,keV)
band, in both soft and medium ($1-2$\,keV) bands, or only in the medium band.
\label{tbl:peaks}}
\begin{tabular}{|l|ccc|}
\tableline
OBSID & $0.4-1$\,keV & both $0.4-1$\,keV & $1-2$\,keV \\
      &  alone       & \& $1-2$\,keV     & alone \\
\tableline
0670130201 & 7 & 9 & 3 \\
0670130301 & 7 & 7 & 5 \\
0670130401 & 8 & 4 & 5 \\
0670130501 & 13 & 8 & 4 \\
0670130601 & 5 & 9 & 3 \\
0670130701 & 0 & 11 & 4 \\
0670130801 & 6 & 8 & 2 \\
0670130901 & 6 & 9 & 3 \\
\tableline
\end{tabular}
\end{center}
\end{table*}

\subsection{The nature of the delayed excess}
The hard excess that is shown in Fig.\,\ref{fig-shapes} has been obtained by averaging many peaks together. The resulting mean shapes for the hard excess could be caused by all peaks having a 
temporally broad, delayed signal with a hard spectrum.
Alternatively, it might be that an individual primary flare tends to be followed by a narrow, secondary peak at a random time offset $\tau$, where $\tau$ follows a probability distribution such that when averaged, we obtain the broad shapes seen. If the latter explanation were true, then we would expect that the excess in any given time bin would be dominated by a few individual peaks. To test for this, we calculated a weighted median of the individual peaks for each time bin, rather than a weighted mean. If the distribution of offset secondary peaks were the explanation, then we would expect the weighted median to fall below the weighted mean. In fact we find that the weighted median agrees with the mean estimate within the uncertainties, indicating that the delayed excess is indeed a smooth continuous function that is consistent between all flares.

\section{Discussion \label{sec:discussion}}

\subsection{Peak shapes and power spectra \label{discussion:psd}}
The peak shapes shown in Fig.\,\ref{fig-shapes} have several common features. The soft-band peak is broadly symmetric. The reference time $t=0$ is defined by this soft-band peak center. The hard-band peaks within $256$\,s of the soft band, and is highly asymmetric, exhibiting an excess at later times. The nature of this excess varies across AGN, being most prominent and extensive in \object{Ark~564}.

The power spectra from the peak shapes (Fig.\,\ref{fig-ps}) are in good agreement with those from the full time series in \object{NGC~4051} and at low frequencies in the other AGN, but lack high-frequency power in \object{Ark~564} and \object{1H~0707--495}. The break frequency in the soft band appears to be the same for the peak shape as for the full series, although the power at frequencies above the break appears to be suppressed in the peak shapes. Three possible explanations of this discrepancy are:
\newcounter{counter2}
\begin{list}{\arabic{counter2}.~}{ \usecounter{counter2}
		 \setlength{\itemsep}{1pt}
     \setlength{\parsep}{1pt}
     \setlength{\topsep}{0ex}
     \setlength{\partopsep}{0pt}
     \setlength{\leftmargin}{1.em}
     \setlength{\labelwidth}{2.em}
     \setlength{\labelsep}{0.em} }
\item There is an uncertainty in the position of the peaks detected in the smoothed time series. This uncertainty broadens the peak shapes, suppressing high frequency power.
\item There might be additional high-frequency variations in the full time series, not captured
by the peak selection.
\item The existence of the rms-flux relation \citep[e.g.][]{mchardy10a} may cause aliasing of modes such that high-frequency
power is boosted with respect to that expected in the simple `moving average' model
\footnote{
The rms-flux relation may be reproduced mathematically by taking a Gaussian time series and 
multiplying by a sinusoidal function in the time domain.  The Fourier transform of the resulting
time series is then expected to be the
convolution of the Fourier transform of the Gaussian time series with the Fourier transform
of the multiplying function.  That convolution aliases power to high frequencies in the powerspectrum.
}.  
\end{list}
However, the good agreement between the PSDs at the frequencies where the positive lags are seen
in all three AGN indicates that the peaks analysis procedure fully captures the timing information
at those frequencies.

When we examine the lag spectra of the transfer functions we find that the soft band lag 
is consistent with zero (although we note that, had we chosen to smooth with an asymmetric smoothing function, the resulting peaks
would have had a non-zero lag at some frequencies).
Thus, the lag on the cross spectrum is primarily due to the shape of the hard band transfer function. 
We particularly note the sharp negative lag feature at $4\times 10^{-4}$\,Hz in
\object{Ark~564} which appears in both the Fourier lag spectrum and the PSD of the
hard-band peak, demonstrating the point of \citet{miller10b} and \citet{miller11a}
that such negative features arise naturally from the effects of Fourier analysis
of reverberation-like transfer functions (section\,\ref{neglags}). A straightforward way to obtain such a negative lag is to have a gap in the transfer function between the `direct' (zero-lag) and `reverberation' (delayed) signals \citep{miller10b}, and this seems to be the case in Ark~564.  However, in other AGN, such a gap could be masked by convolution with a broad flare temporal response, as appears to be the case in 1H0707-495.  If the flare temporal response is symmetric, the convolution effect does not change the cross-spectrum phases, and hence the time lags in the lag spectrum, but can reduce the appearance of a gap in the time domain.

\subsection{Possible explanations of the observed temporal behavior}

\subsubsection{Intrinsic flare temporal structure}
We now consider possible explanations for the lag spectra that have been discussed in the literature. 
At present, we have little idea of the mechanism that creates the flares and 
the general time series variations.  It is possible therefore that any temporal structure
may be intrinsic to some complex high-energy process and not a consequence of 
propagation of either photons or sound through an extended distribution. For example, solar flares
are generated by a rather poorly-understood mechanism, and also show features such as
systematic spectral variations in hard X-rays \citep[e.g.][]{krucker02a}, although we note that
the solar X-ray spectra and spectral variations are complex and 
differ substantially from those observed in AGN.
Until the primary flare mechanism has been definitively established in AGN, we would never be
able to eliminate this possibility.  However, we note that the timescale of the lags in
\object{Ark~564} are substantially longer than the duration of the primary flare, and
it does seem likely that the delayed, hard-spectrum excess seen in Ark~564 arises 
through secondary reprocessing rather than through the primary flare process.

The durations of the primary flares measured in section\,\ref{sec:peakselection} of
$370-500$\,s allow upper limits to be deduced for the sizes of the regions producing those flares.  
The black hole masses are uncertain (section\,\ref{introduction}), but the flare durations imply 
light crossing-times $d \la 50$\,$r_{\rm g}$.

\subsubsection{Inverse Compton upscattering time delays}
Our analysis disfavors models 
that explain the lag as being due to inverse Compton upscattering of photons, 
in which the hard band variations are delayed by the time taken to upscatter
photons from the soft band. 
We find no sign of such a delay of the hard-band peak with respect to the soft-band
peak in the peak shapes we extract: the positive lags appear to be entirely due to
the additional hard-spectrum delayed component alone.  
The primary flares may be produced by inverse Compton upscattering, if the 
upscattering timescale is shorter than the temporal resolution of our analysis.  The
hard-band excess has timescales a factor $\ga 10$ longer, not consistent with
arising as inverse Compton upscattering delay in the flare generation region.

\subsubsection{Positive lags from propagating fluctuations}
\citet{arevalo06b} have proposed that the low-frequency, positive lags in AGN arise from fluctuations in the accretion disk that propagate radially inwards, with
an assumption that a fluctuation at large radius results in a soft X-ray emission spectrum, whereas a fluctuation at smaller radius results
in a harder X-ray spectrum. This assumption is made because the accretion disk at small radii is hotter than at larger radii in the standard thin-disk model \citep{shakura73a}.  The viability of such a model in AGN is open to question. It requires inward-propagating modes to dominate the variability of X-rays. Also an AGN accretion disk itself is, unlike that of a galactic black hole candidate, insufficiently hot to emit significant hard X-rays. Instead the disk is required to act as a source of seed UV emission to an inverse Compton upscattering process, and the process is thought to operate in the regime where the spectral index of the upscattered radiation is largely independent of the seed photon distribution \citep{titarchuk97a}. This potential issue can be gotten around by allowing radiation from different parts of the disk to illuminate different Comptonizing regions with different properties. 

This model is also disfavored by the measurements of Ark~564 presented here:  the flaring behavior shows sharp peaks that are temporally coincident at all photon energies, accompanied by a delayed, hard-spectrum signal, whereas in the simplest `propagating fluctuations' model we would expect to see a broad temporal profile in which the spectrum hardened with time. Such a model would be consistent with the other two AGN, whose delayed signals 
appear as asymmetries in the peak shape rather than in the form of a distinct component, but if we
wish to find a single explanation for the lags in all AGN, the observation of time-coincident
peaks accompanied by a distinct, hard, secondary maximum in Ark~564 remains challenging.  

To explain this, a more sophisticated model of `propagating fluctuations' might suppose that a flare
may be produced at some location, either on the accretion disk or in the corona, which is
directly observed at all energies, but that a disturbance then propagates into a hard-spectrum
region such that a delayed,
hard signal is seen.  In Ark~564 there is a distinct gap in the transfer function after
the primary peak which does not have an obvious explanation in this scenario.  
It is worth noting that the gap is responsible for the sharp negative lag in the lag spectrum
of Ark~564, and is a significant feature to be explained (see section\,\ref{neglags}).

Such a model might, however, be able to explain the observation of section\,\ref{sec:waveband},
that the delayed signal is stronger for flares that have soft spectra, and absent for those
flares that have hard spectra.  A possible explanation is that the soft flares create 
propagating fluctuations that travel into the hard-emitting region, whereas hard flares might
be created directly in that region.

\subsubsection{X-ray reverberation}\label{sec:xray_reverb}
The final model considered here is the reverberation explanation of \citet{miller10b, miller10a} and \citet{miller11a}.  In those papers, it was shown how a long-timescale delay, such as might be produced by reverberation from gas a few tens to thousands of gravitational radii from the X-ray source, could naturally produce low-frequency positive lags, and also potentially generate high-frequency negative lags as a result of the shape of the transfer functions in the two bands being cross-correlated \citep{miller10b, miller11a}.  Such models are consistent with the analysis presented here, and in particular with the delayed, hard excesses that are seen in all three AGN,
but particularly prominently in \object{Ark~564}.  

\begin{figure}[!h]
\epsscale{1.}
\plotone{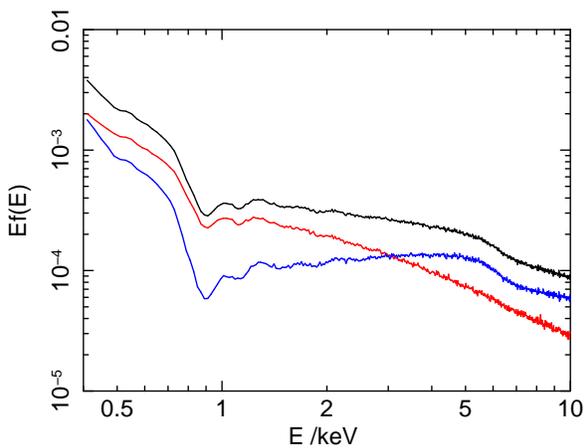}
\caption{
Example model spectrum from the uniform spherical scattering model described in the text.  The upper black
curves shows the total spectrum received by the observer.  The directly received component (red curve)
and delayed, scattered component (blue curve) are also shown.
\label{fig:toymodel}
}
\end{figure}

The fraction of hard-band photons in the delayed excess to photons in the primary peak is very high, 
with a mean value of
$1.65 \pm 0.2$, but apparently being as large as $2.8 \pm 0.6$ in the first half of the observations
(section\,\ref{sec:shape-ark-byhalf}).  Within the reverberation hypothesis, this could arise
if the observer's sight-line to the primary flare site were through material that is moderately
optically-thick to Compton scattering, so that most photons from the flare are scattered and
appear in the delayed excess.  The dependence on flare hardness could arise naturally in
ionized gas as, for a wide range of values of the ionization parameter, the photoelectric opacity in the
$1-2$\,keV band can be larger than that in the $0.4-1$\,keV band.  Thus in the first half of the observations,
we could be looking through a higher optical depth of scattering, absorbing gas than in the later
half, leading both to softer flares around $0.4-2$\,keV and to a larger delayed excess in the $4-7.5$\,keV band.

As a `toy model' example of this, we have calculated the spectrum expected from a uniform sphere of gas 
near the surface of which is embedded a point source of X-rays with a powerlaw continuum.  Spectra
have been calculated by a 3D Monte-Carlo radiative transfer algorithm with full treatment of Compton scattering
by cold, free electrons, but assuming that photoelectric opacity is provided by gas with an invariant
ionization parameter, where opacities have been obtained from {\sc xstar}
\citep{kallman04a}.  No thermal or ionization balance calculations have been made inside the simulation and line
and radiative recombination continuum emission has been ignored.  
The effects of thermal and macroscopic Doppler shifts have been approximated by smoothing with a Gaussian velocity dispersion $\sigma=0.05$c, appropriate for material at radii $r\sim 100$\,r$_g$.
This simple model is not expected to be able to fit data in any detail (and we stress that this does not represent a spectral fit to any of the sources presented in this paper), but it does illustrate the
principal features of having a high fraction of scattered light with a hard spectrum.  
Fig.\,\ref{fig:toymodel} shows an example for ionization parameter $\log_{10}\xi = 2.1$, 
electron column density $N_{\rm e} = 4\times 10^{23}$\,cm$^{-2}$ and mean absorber
hydrogen column density $N_{\rm H} = 4 \times 10^{22}$\,cm$^{-2}$ in front of the X-ray source
(in practice achievable by having a two-phase medium of hot, free electrons within within are
embedded denser clumps of lower ionization gas).  The source
has a powerlaw index $\Gamma = 2.5$. The depth of the source within the sphere is 1/50 the sphere radius.  This
configuration gives a ratio of scattered to direct light within the $4-7.5$\,keV band of 1.65 as in
\object{Ark~564}.

An alternative explanation for the high amplitude of the delayed excess might be that the reverberating
region is itself sufficiently hot that additional photons are inverse Compton upscattered into the
hard energy band by the reverberating gas.  Such models may also naturally produce a high fraction
of X-rays in the delayed excess, as there is a large source of soft photons that may be suitably
upscattered.  We would expect the resulting spectra to be largely featureless, generally consistent
with the observations of \object{Ark~564}.
Models that seek to link the transfer function behavior and the spectral properties of Ark~564
will be discussed by Giustini et~al. (in preparation).

The scale-free example model presented above may be scaled in size to produce the observed time lags.  However, we do not expect that a simple uniform sphere will reproduce the detailed shape of the transfer function, for which more extensive radiative transfer modeling of clumpy, anisotropic distributions will be required.

The differing spectra of flares inferred in section \ref{sec:waveband} need not imply rapid variations in ionization parameter or optical depth. Instead, it could be that flares are emitted from multiple regions, each with their own value of optical depth. The longer timescale changes described in sections \ref{sec:nonstationary} \& \ref{sec:shape-ark-byhalf} are of order weeks (the time between observations).

\subsubsection{Negative lags \label{neglags}}
Negative lags are seen in the lag spectra of \object{1H~0707--495} and
\object{Ark~564}, but not in \object{NGC~4051}.  In \object{Ark~564}
there is a sharp negative lag which is directly linked to the gap in
the transfer function between the primary peak and the delayed signal
(see the discussion of transfer functions by \citealt{miller10b} and
\citealt{miller11a}).  
This seems to confirm one of the basic points
of \citet{miller10b}, that sharp negative oscillations may be caused simply by
the shape of the hard band transfer function, and not by any requirement for a
secondary scattering process in the soft band.

However, the peaks analysis presented here lacks sufficient signal-to-noise to detect
the oscillatory, negative lags found at higher frequencies in the
lag spectra of these two AGN.  Persistent negative lags that extend over a wide range
of frequency are difficult to explain
if it is only the hard band that has an extended transfer function
(see \citealt{miller10b}, \citealt{miller11a} and \citealt{zoghbi11a}) and
it seems that an extended transfer function may be needed also in the soft band
\citep{miller10b}.
In the case of Ark\,564, those high-frequency negative lags are low amplitude,
oscillatory and may also be non-stationary (section\,\ref{sec:nonstationary}).
If it is indeed required, 
our analysis cannot detect such a soft-band transfer function, as it is
limited by the attainable signal-to-noise, but we should nonetheless discuss
the requirements that the possible models would need to satisfy in each case.

The `propagating fluctuations' model does not naturally explain those
negative lags in \object{1H~0707--495}, and \citet{fabian09a} and \citet{zoghbi10a,
zoghbi11a} have proposed that these arise as a separate process, in
which a compact X-ray source, close to both the black hole and the
accretion disk, illuminates the disk and produces X-ray emission. With
a suitably-tuned reflection spectrum, enhanced by gravitational bending of light
near the black hole, it was proposed that this may
result in a delayed reverberation signal in the soft band.  Some of
the difficulties of this model have been discussed by \citet{miller11a} and
are not discussed further here, apart from noting that the model requires 
(i) a very high abundance of iron (at least a factor 11 super-solar, \citealt{zoghbi10a}), 
(ii) the X-ray source to be located within a few gravitational radii of the event
horizon of a maximally-rotating black hole \citep[see also][]{chainakun12a} and 
(iii) the AGN to be significantly super-Eddington\footnote{
The black hole mass for \object{1H~0707--495} of $2\times 10^6$\,M$_\odot$ assumed
by \citet{zoghbi10a} led to those authors associating a 30\,s negative lag
with reflection from a radius 2\,r$_g$.  Given the inferred bolometric luminosity
\citep{leighly04a} such a black hole would be about a factor 5 super-Eddington.
If the black hole mass were higher, the radius of the reflecting region would be
correspondingly smaller, and becomes implausibly close to the event horizon
for black hole masses $\ga 10^7$\,M$_\odot$.
We note that the black hole mass of \object{1H~0707--495} 
has not yet been determined by optical reverberation
mapping, at the time of writing.
}. 
Thus an inverse Compton upscattering source, required to produce the powerlaw hard X-ray spectrum
and postulated by \citet{arevalo06a}, would need
to be extremely close to the black hole and to produce short-timescale
flares with long-timescale variable emission
that is both delayed and significantly harder than the prompt flare
spectrum.  

As \object{Ark~564} also shows similar negative lags to \object{1H~0707--495}, the complex
spectral model invoked for \object{1H~0707--495} should also be required for
\object{Ark~564}, yet the 0.4--10\,keV spectrum of \object{Ark~564} appears 
to differ substantially from that of \object{1H~0707--495}
(see Giustini et~al., in preparation).

\citet{miller10b} proposed that, 
in the reverberation model, some soft-band X-rays scattered with short delays
in a continuous medium could reach the observer,
but that long-delay scattered soft X-rays would be absorbed within the scattering medium
owing to the longer path-length traversed,
leading to a soft-band transfer function that was extended over short timescales only.  
While such a scenario is possible,
photoelectric absorption can only remove photons from an X-ray energy band, implying that
the soft-band transfer function should always be lower in amplitude than the hard band
transfer function, 
so this mechanism might require fine-tuning to obtain an appropriate cross-correlation signal.

If we follow \citet{fabian09a} and \citet{zoghbi10a, zoghbi11a}, and
suppose that there are two separate mechanisms that operate at low and high frequencies,
an alternative possibility is that the short timescale ($\la 30$\,s) negative lags are
actually a feature of the emitting region itself.  Lags on that timescale are 
significantly shorter than
the duration of the flares ($\sim 400-500$\,s), 
and so are more plausibly associated with the primary process
than are the ks-timescale delayed signals seen at low frequencies.   
If the flare-emitting region were slightly larger at soft energies than at hard energies,
soft-band flares would have PSDs that cut off at lower frequency than hard-band flares
(or, equivalently, slightly broader peaks in the peaks analysis),
as observed, and would have variations delayed by time $\Delta \tau \simeq \Delta r/c$, where
$\Delta r$ is the difference in apparent size of the flare region between the soft and hard
X-ray bands.  Such a size differential could arise either from the effects of photoelectric
opacity or from the energy-dependent diffusion of electrons from a flare acceleration region,
given that radiative lifetimes due to inverse Compton scattering increase with decreasing electron
energy.

\section{Conclusions}
This paper has explored the origin of the PSDs and time delays in three,
highly X-ray variable, narrow-line Seyfert\,I AGN: 
\object{NGC~4051}, \object{1H~0707--495} and \object{Ark~564},
using previously-published data for the first two and a new long 
{\em XMM-Newton} observation of \object{Ark~564}.  The lag spectrum
of \object{Ark~564} shows strong positive lags (hard photons lagging soft photons)
for Fourier modes with low 
frequencies, consistent with previous work
\citep{arevalo06a, mchardy07a}. 
It also shows a sharp negative lag at $4\times 10^{-4}$\,Hz and oscillatory, slightly
negative lags at higher frequencies.  The PSD of \object{Ark~564}
has a complex shape that includes non-smooth structure, but which cannot be
clearly labeled as a quasi-periodic oscillation.
We have also found evidence for non-stationarity of the PSD and lag spectra
of \object{Ark~564} on timescales of weeks.

To better understand the lag spectra of these AGN,
we have developed a method for estimating the transfer function of AGN
X-ray time series in individual bands of photon energy, which allows
us to investigate physical models of time lags without the limitation
of using cross-correlation between bands.  
By modeling the AGN variations as a `moving average' time series process, 
the method obtains the mean shape
of emission peaks in the light curve, after a nested 
median filtering process to remove long-timescale
variations has been applied.  All three AGN show an excess of
delayed emission in hard X-rays, but not soft X-rays,
on ks timescales following an emission flare. 
The delayed emission detected in the peaks analysis is able to reproduce
the lags between hard and soft photons at low frequencies that are deduced from
Fourier analysis of the time series.  The sharp negative lag feature 
in \object{Ark~564} is
seen to arise from the shape of the transfer function in the hard band,
and not from time delays in soft X-rays, as discussed in the context of
models of X-ray reverberation by \citet{miller10a} and \citet{miller11a}.

In \object{Ark~564}
the delayed hard excess is particularly prominent and appears to be a distinct
secondary reprocessing of the flare photons, as expected from reverberation
caused by Compton scattering of photons by circumnuclear material.  A high
optical depth to Compton scattering is implied by the high fraction of photons
in the delayed excess.  The amplitude of the excess appears to vary both with
time and with the hardness of the selected peaks, implying that we are observing
a complex, rapidly-varying region of this AGN.
The temporal coincidence in all three AGN
of the primary peaks of emission between hard and soft
photons is inconsistent with the time lags being caused by inverse Compton upscattering
delays, and the shape of the \object{Ark~564} excess is difficult to reconcile
with a model in which time lags are caused by fluctuations propagating
across an accretion disk.
Given the estimated black hole masses in these AGN,
the transfer functions we estimate are broadly
consistent with reverberation from material 10s to 100s of
gravitational radii from the central source and provide further evidence for
the existence of a substantial covering fraction of scattering material in the
central regions of type\,I AGN.

\acknowledgments

EL acknowledges support from an STFC postgraduate studentship.
LM acknowledges support from STFC grant ST/H002456/1. TJT and MG 
acknowledge support from NASA grant NNX08AJ41G.

Observations of \object{Ark~564} and \object{1H~0707--495} were obtained with \facility{{\em XMM-Newton}}, 
an ESA science mission
with instruments and contributions directly funded by
ESA Member States and NASA.
Analysis of \object{NGC~4051} made use of data obtained from the \facility{{\em Suzaku}} 
satellite, a collaborative mission between the space agencies of Japan (JAXA) and the USA (NASA).

\bibliographystyle{apj}
\bibliography{xray}

\end{document}